\documentstyle[twoside,fleqn,espcrc2,epsf]{article}

\newcommand{\half}{\mbox{\small $\frac{1}{2}$}}          

\def\lsim{\mathrel{\rlap{\lower4pt\hbox{\hskip1pt$\sim$}}
    \raise1pt\hbox{$<$}}}                
\def\gsim{\mathrel{\rlap{\lower4pt\hbox{\hskip1pt$\sim$}}
    \raise1pt\hbox{$>$}}}                

\title{
       \vspace{-3.65cm}                                     %
       {\normalsize DESY 02-148}      \\[-0.2cm]            
       {\normalsize Edinburgh 2002/08}\\[-0.2cm]            
       {\normalsize LU-ITP 2002/017}\\[-0.2cm]              
       {\normalsize September 2002}  \\                     
       \vspace{1.78cm}                                      
       A non-perturbative determination of $Z_V$ and $b_V$ for $O(a)$
       improved quenched and unquenched Wilson fermions%
            \thanks{Talk given by R. Horsley at Lat02,
                    Boston, U.S.A.}}                        

\author{T. Bakeyev%
           \address{Joint Institute for Nuclear Research,
                    141980 Dubna, Russia},
        M. G\"ockeler%
           \address{Institut f\"ur Theoretische Physik, Universit\"at
                    Leipzig, D-04109 Leipzig, Germany}
                       \hspace{-0.25cm} $^,$\hspace{-0.15cm}
           \address{Institut f\"ur Theoretische Physik, Universit\"at
                    Regensburg, D-93040 Regensburg, Germany},
        R. Horsley%
           \address{School of Physics,
                    University of Edinburgh, Edinburgh EH9 3JZ, U.K.},
        D. Pleiter%
           \address{John von Neumann Institute NIC / DESY Zeuthen,
                    D-15738 Zeuthen, Germany},
        P.~E.~L. Rakow%
           $^{\rm c}$,
        A. Sch\"afer%
           $^{\rm c}$,
        G. Schierholz%
           $^{\rm e,}$%
           \address{Deutsches Elektronen-Synchrotron DESY,
                    D-22603 Hamburg, Germany}
        and
        H. St\"uben%
           \address{Konrad-Zuse-Zentrum f\"ur Informationstechnik Berlin,
                    D-14195 Berlin, Germany}
        -- {\it QCDSF} Collaboration }

\begin{document}

\begin{abstract}
By considering the local vector current between nucleon states
and imposing charge conservation we determine, for $O(a)$ improved
Wilson fermions, its renormalisation constant and quark mass improvement
coefficient. The computation is performed for both
quenched and two flavour unquenched fermions.
\end{abstract}

\maketitle

\setcounter{footnote}{0}


\section{INTRODUCTION}

Due to the presence of the `Wilson term' in the lattice
fermion action for Wilson fermions
the discretisation errors are $O(a)$. As the gluon part of the action
(sum of plaquettes) has only $O(a^2)$ errors, it is also desirable to achieve
this for the fermion action. The Symanzik programme%
\footnote{For a recent review see, for example, \cite{luscher98a}.}
allows a systematic reduction of errors to $O(a^2)$ upon including
additional higher dimensional operators. The (on-shell) action is improved
with a suitably tuned `clover' term. However it is also necessary
to improve each operator separately. Much work has been devoted
to this topic in recent years; here we shall just concentrate
on the local vector current: $V^{(q)}_\mu = \overline{q}\gamma_\mu q$. 
In this case just two additional operators $am_q V^{(q)}_\mu$
and $\half ia\partial_\lambda T^{(q)}_{\mu\lambda}$ are required
giving the $O(a)$ improved and renormalised vector current as
\begin{eqnarray}
  {\cal V}_\mu^{(q)R} = Z_V ( 1 + am_q b_V )
      ( V^{(q)}_\mu + \half ia c_V \partial_\nu T^{(q)}_{\mu\nu})
                                                \nonumber
\end{eqnarray}
with $T^{(q)}_{\mu\nu} = \overline{q} \sigma_{\mu\nu} q$. The second
improvement operator only has an effect in {\it non-forward} matrix elements
and will not be considered further here. The renormalisation constant
$Z_V$ and improvement coefficient $b_V$ are functions of the coupling
constant $g_0$ and perturbatively we have,
\cite{gabrielli91a}, to one loop (independently of the presence of fermions),
\begin{eqnarray}
   Z_V(g_0) &=& 1 - 0.12943 g_0^2 + \ldots ,
                                               \nonumber \\
   b_V(g_0) &=& 1 + 0.15323 g_0^2 + \ldots ,
                                               \nonumber
\end{eqnarray}
but in presently accessible regions of $\beta \equiv 6/g_0^2$ 
there may be considerable deviations.


\section{THE CONSERVED CURRENT}

\begin{figure}[htb]
   \vspace*{0.05in}
   \epsfxsize=7.25cm \epsfbox{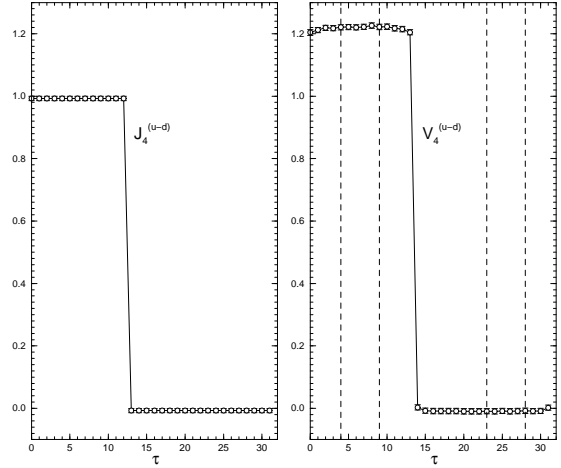}
   \vspace*{-0.30in}
   \caption{\footnotesize{\it $R(J_4^{(u-d)})$ and $R(V_4^{(u-d)})$
            plotted against the operator position $\tau$ for
            the quenched ($n_f=0$) data set $\beta = 6.0$,
            $\kappa = 0.1342$ on a $N_S^3\times N_T = 16^3\times 32$
            lattice with $t=13$. Typical fit intervals for
            $V_4^{(u-d)}$ are given by the pairs of vertically
            dashed lines.}}
   \vspace*{-0.30in}
   \label{fig_Rratio_bare_b6p00kp1342_standard}
\end{figure}

There is an exact symmetry of the action
$q \to e^{-i\alpha}q$, $\overline{q} \to e^{i\alpha} \overline{q}$
giving via the Noether theorem the Ward identity (WI)
\begin{eqnarray}
   \langle \Omega \overline{\Delta}_\mu J^{(q)}_\mu \rangle =
        \left\langle {\partial\Omega \over \partial q}q \right\rangle + 
        \left\langle \overline{q}{\partial\Omega \over \partial \overline{q}} 
        \right\rangle ,
                                               \nonumber
\end{eqnarray}
where $\Omega$ is an arbitrary operator,
$\overline{\Delta}_\mu$ is the lattice backward derivative,
and $J^{(q)}_\mu$ the exactly conserved vector current%
\footnote{$O(a)$ improvement for non-forward matrix elements requires,
as for the local vector current, an additional operator
$\half ia \half [ \partial_\nu T^{(q)}_{\mu\nu}(x) +
                              \partial_\nu T^{(q)}_{\mu\nu}(x+\hat{\mu}) ]$.}
($CVC$),
\begin{eqnarray}
   \lefteqn{J^{(q)}_{\mu}( x + \half \hat{\mu}) =}
      & &
                                                \nonumber  \\
      & & \hspace*{-0.275in}
          \half \left[ \overline{q}_x(\gamma_\mu -1)U_\mu(x)q_{x+\hat{\mu}} -
                \overline{q}_{x+\hat{\mu}}(\gamma_\mu +1)U_\mu(x)^\dagger q_x
                \right]
                                               \nonumber
\end{eqnarray}
Roughly speaking the RHS of this equation counts the number of $q$ and
$\overline{q}$ in operator $\Omega$.
For our numerical results we take
$\Omega \to B \overline{B}$ where $B$ is the (unpolarised)
nucleon operator to give, upon solving the WI equation,
\begin{eqnarray}
   R(J^{(u-d)}_4)
                 &=& { \langle B(t) J^{(u-d)}_4(\tau) \overline{B}(0)
                       \rangle
                     \over \langle B(t) \overline{B}(0) \rangle}
                                                \nonumber \\
                 &=& \left\{
                     \begin{array}{ll}
                        c_1^{(u-d)} & 0 < \tau < t     \\  
                        c_2^{(u-d)} & t < \tau < N_T-1
                     \end{array} \right. 
                                               \nonumber
\end{eqnarray}
($J_4^{(u-d)} \equiv J_4^{(u)} - J_4^{(d)}$)
where $c_i^{(u-d)}$ are constants and with
{\it jump}, $\Delta R(J_4^{(u-d)}) \equiv c_1^{(u-d)} - c_2^{(u-d)} = 1$,
ie charge conservation.
The numerical advantage of considering $u - d$ is that the hard to
compute quark line disconnected terms cancel.
(For the $CVC$ this term vanishes though.)
In Fig.~\ref{fig_Rratio_bare_b6p00kp1342_standard} we show this ratio
for the conserved vector current.
A very good signal is observed (indeed the result should be true to machine
accuracy).


\section{THE LOCAL CURRENT}

\begin{figure}[htb]
   \vspace*{-0.05in}
   \epsfxsize=7.00cm
      \epsfbox{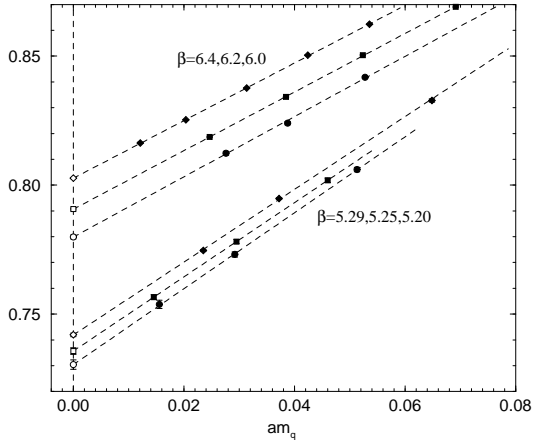}
   \vspace*{-0.30in}
   \caption{\footnotesize{\it $\Delta R(V_4^{(u-d)})$
            for quenched configurations for $\beta = 6.4$, $6.2$
            and $6.0$ (upper set of curves, top to bottom respectively)
            and for the unquenched configurations for $\beta = 5.29$,
            $5.25$ and $5.20$ (lower set of curves).}}
   \vspace*{-0.30in}
   \label{fig_V_0_1u-1d.p0_020509_1244_lat02_expt_wrtup}
\end{figure}

The local vector current ($LVC$) is not conserved on the lattice
and so we do not expect the jump to be equal to one.
This is shown in the RH picture in
Fig.~\ref{fig_Rratio_bare_b6p00kp1342_standard}. We now define
the renormalisation constants  ($Z_V$, $b_V$) by demanding that
the renormalised local current has the same behaviour
as the conserved current, so that
\begin{eqnarray}
   Z_V ( 1 + am_q b_V) =
        \left[ \Delta R( V_4^{(u-d)} ) \right]^{-1} .
                                                \nonumber
\end{eqnarray}
Thus upon plotting the data, the intercept gives $Z_V$ and the gradient
$Z_Vb_V$. ($am_q = \half ( 1/\kappa - 1/\kappa_c )$ and $\kappa_c(g_0)$ is 
estimated from $r_0 m_q \propto (r_0m_{ps})^2$.) In
Fig.~\ref{fig_V_0_1u-1d.p0_020509_1244_lat02_expt_wrtup} we show
quenched results from which the intercept and gradient can be found.
Other alternative non-perturbative determinations have been given by the
ALPHA collaboration, \cite{luscher96a}, using the Schr\"odinger functional,
the LANL collaboration, \cite{bhattachary01a}
using other Ward identities and Martinelli et al., \cite{martinelli94a}
by `mimicking' perturbation theory.

As well as quenched data sets ($n_f=0$), in collaboration with UKQCD,
we have also generated unquenched data sets.
In this study we use the configurations
with parameters given in Table~\ref{table_unquenched_data_sets}.
\begin{small}
\begin{table}[h]
   \vspace*{-0.40in}
   \begin{center}
      \begin{tabular}{||l|l|l|l|l||}
         \hline
          $\beta$    & $\kappa_{sea}$& Volume& Trajs.    &  Group \\
         \hline
 5.20 & 0.1342 & $16^3\times 32$ & 5000 & QCDSF\\
 5.20 & 0.1350 & $16^3\times 32$ & 8000 & UKQCD\\
 5.20 & 0.1355 & $16^3\times 32$ & 8000 & UKQCD\\
         \hline
 5.25 & 0.1346 & $16^3\times 32$ & 2000 & QCDSF\\
 5.25 & 0.1352 & $16^3\times 32$ & 8000 & UKQCD\\
 5.25 & 0.13575& $24^3\times 48$ & 1000 & QCDSF\\
         \hline 
 5.29 & 0.1340 & $16^3\times 32$ & 4000 & UKQCD\\
 5.29 & 0.1350 & $16^3\times 32$ & 5000 & QCDSF\\
 5.29 & 0.1355 & $24^3\times 48$ & 2000 & QCDSF\\
         \hline
      \end{tabular}
   \end{center}
   \caption{\footnotesize{\it Data sets used in the unquenched,
            $n_f=2$, simulations.}}
   \vspace*{-0.35in}
   \label{table_unquenched_data_sets}
\end{table}
\end{small}

\begin{figure}[htb]
   \vspace*{-0.12in}
   \epsfxsize=7.00cm 
      \epsfbox{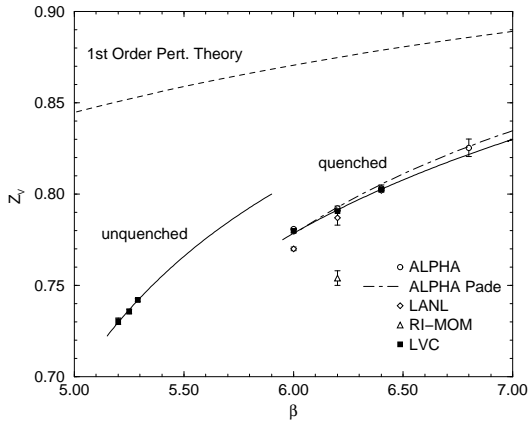}
   \vspace*{-0.32in}
   \caption{\footnotesize{\it $Z_V$ ($LVC$, filled squares) determined in
            this work for quenched and unquenched $O(a)$ improved
            fermions. For the quenched case a comparison is made
            with ALPHA, \protect{\cite{luscher96a}},
            LANL, \protect{\cite{bhattachary01a}} and
            RI-MOM, \protect{\cite{martinelli94a}}. Pad{\'e}
            fits are also given for our and the ALPHA results.}}
   \vspace*{-0.30in}
   \label{fig_zv+bv_zv_beta_magic_qu+dyn_lat02_wrtup}
\end{figure}

\begin{figure}[htb]
   \epsfxsize=7.25cm 
      \epsfbox{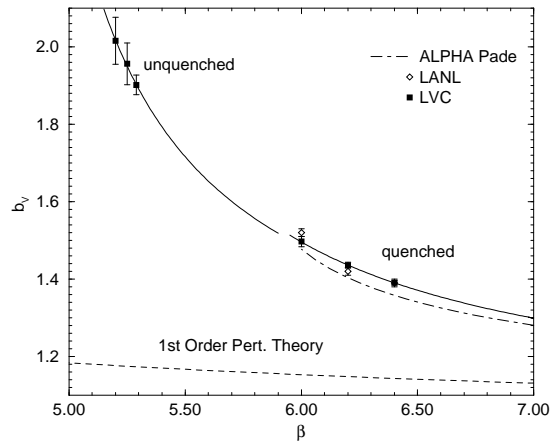}
   \vspace*{-0.30in}
   \caption{\footnotesize{\it $b_V$ ($LVC$, filled squares) determined in
            this work again for both quenched and unquenched 
            $O(a)$ improved fermions. Also shown is the Pad{\`e} fit
            from ALPHA, \protect{\cite{luscher96a}}, and the LANL,
            \protect{\cite{bhattachary01a}} results for quenched
            fermions.}}
   \vspace*{-0.25in}
   \label{fig_zv+bv_bv_beta_magic_qu+dyn_lat02_wrtup}
\end{figure}

We now present our results.
In Fig.~\ref{fig_zv+bv_zv_beta_magic_qu+dyn_lat02_wrtup}
we show $Z_V$ and in Fig.~\ref{fig_zv+bv_bv_beta_magic_qu+dyn_lat02_wrtup},
$b_V$. For quenched fermions good agreement with other methods is found.


\section{CONCLUSIONS}

The method described here reproduces the results of other approaches for
$O(a)$ improved quenched fermions. (But one needs to remember that
$Z_V$ definitions can vary by $O(a^2)$ while $b_V$ definitions
may vary by $O(a)$.) For $O(a)$ improved unquenched fermions $Z_V$ is
{\it smaller} and $b_V$ {\it larger} than for quenched fermions
at the same lattice spacing (roughly $a_{n_f=2}(5.25) \sim a_{n_f=0}(6.0)$).
Further details and final results will appear in \cite{gockeler02b}.


\section*{ACKNOWLEDGEMENTS}

The numerical calculations were performed on the Hitachi {\it SR8000} at
LRZ (Munich), the APE100, APEmille at NIC (Zeuthen) and
the Cray {\it T3E}s at EPCC (Edinburgh), NIC (J\"ulich), ZIB (Berlin).
We thank the {\it UKQCD} Collaboration
for use of their unquenched configurations.
TB acknowledges support from INTAS grant 00-00111.
This work is supported by the European Community's Human potential
programme under HPRN-CT-2000-00145 Hadrons/LatticeQCD.



\end{document}